\documentclass[aps,prl,superscriptaddress,twocolumn,showpacs]{revtex4-1}
\usepackage[pdftex]{graphicx}
\usepackage{placeins}
\usepackage{amsmath}

\begin{document}


\title{Coupled pair of one and two dimensional magneto-plasmons on electrons on helium.}



\author{A.D. Chepelianskii}
\affiliation{LPS, Univ. Paris-Saclay, CNRS, UMR 8502, F-91405, Orsay, France}
\author{ D. Papoular}
\affiliation{ LPTM, UMR 8089 CNRS \& Univ. Cergy–Pontoise, France }
\author{ D. Konstantinov}
\affiliation{Okinawa Institute of Science and Technology (OIST) Graduate University, Onna, Okinawa 904-0412, Japan}

\author{H. Bouchiat}
\affiliation{LPS, Univ. Paris-Saclay, CNRS, UMR 8502, F-91405, Orsay, France}

\author{K. Kono}
\affiliation{ICST, NCTU, 1001 Ta Hsueh Rd., Hsinchu 300, Taiwan }
%
%
%


\date{\today}

\begin{abstract}
Electrons on the liquid helium surface form an extremely clean two dimensional system where different plasmon-excitations can coexist. Under a magnetic field time reversal symmetry is broken and all the bulk magneto-plasmons become gapped at frequencies below cyclotron resonance while chiral one dimensional edge magneto-plasmons appear at the system perimeter. We  theoretically show  that  the presence  of  a homogeneous  density
gradient in the  electron gas leads to the formation  of a delocalized
magneto-plasmon  mode  in  the  same frequency  range  as  the  lowest
frequency  edge-magnetoplasmon mode.   We  experimentally confirm  its
existence by measuring the corresponding resonance peak in frequency dependence of the admittance of the electron gas. This allows to realize a prototype system to investigate the coupling between a chiral one-dimensional mode and a single delocalized bulk mode. Such a model system can be important for the understanding of transport properties of topological materials where states of different dimensionality can coexist. 
\end{abstract}


\maketitle

The recent discovery of topological states of matter has lead to striking predictions of topological surfaces and edge states \cite{Kane2005, Bernevig2006, Konig2007, Fu2007, Murakami2007, Hsieh2008, Konig2007, Chen2009, Taskin2011, Knez2011, Murani2017, Schindler2018}. 
However, it has so far been difficult to realise a system where topological edge states are completely decoupled from remaining bulk 
states or spurious edge states of non-topological origin \cite{Kvon2015,Ensslin2015, Murani2017, Schindler2018}. Thus understanding the interaction between topological edge modes and non-topological bulk 
modes is highly important. Electrons on helium are a high purity two dimensional system where chiral edge magnetoplasmons modes naturally form under a perpendicular magnetic field \cite{Glattli1985,Mast1985,Ivana2013,Volkov1988}, interestingly their topological origin has been recognized only recently \cite{Jin2016,Jin2018,topoPlasmon}. 
Bulk magneto-plasmon in a two dimensional electron gas have a gap at frequencies below the cyclotron resonance, and it is traditionally considered that edge and acoustic magnetoplasmons are the only low frequency plasmon excitations \cite{Volkov1991,Elliott1995,Aleiner1995}. In experiments with electrons on helium the frequency of edge magnetoplasmons (EMP) is typically in the kHz range, while the cyclotron resonance frequency is typically several GHz. We show how a low energy bulk mode can be created inside the bulk-magnetoplasmon gap by an anisotropic gradient of electronic density.
In this letter we will describe this plasma excitation as a magneto-gradient mode.  
We show that the frequency of this magneto-gradient mode can be obtained from an effective Schr\"odinger equation allowing to control the resonance 
frequency through the shape of the electron cloud. This allows to tune this frequency into resonance with the edge magnetoplasmons 
creating a model setting to study the interaction between bulk and topological edge modes. We note that the existence of this low frequency bulk magneto-plasmon can also be important to understand the surprising collective effects that appear in electrons on helium in the microwave induced resistance oscillation regime \cite{miro1,miro2,zrs1,zrs2,miro4,miro5,miro6}: zero-resistance states \cite{Denis1,Denis2}, incompressibility \cite{Alexei2015,Denis3} and self-oscillations \cite{DenisWatanabe} which are not yet understood microscopically.

We first show that the presence of a density gradient can indeed lead to the formation of a low frequency delocalized magneto-plasmon, this may seem counter-intuitive as in an homogeneous system all the bulk magneto-plasmons are gapped with their lowest frequency given by the cyclotron frequency $\omega_c = e B/m$. The equations of motion for magnetoplasmons can be derived from the drift transport equations on the electronic density $n_e = n_{a} + n_{t}$ where we decompose the electronic density into time--averaged ($n_a$) and time--dependent ($n_t$) parts (see a sketch of the cell geometry in Fig.~1). Treating the time-dependent terms as a small perturbation, the linearised transport equations are: 
\begin{align}
\partial_t n_{t} = {\rm div}_{2d} \left[ n_{a} \left( \mu_{xx} \nabla_{2d} V_{t} + \mu_{xy} \mathbf{u}_z \times  \nabla_{2d} V_{t} \right) \right] 
\label{dtne}
\end{align}
Here $V_{t}$ is the time dependent part of the quasi-static electric potential $V = V_{s} + V_{t}$, and $V_s$ is its static part. In deriving this equation we took into account that the electron cloud screens the static-part of the electronic potential which leads to $\nabla V_{s} = 0$, the longitudinal and Hall mobilities are given by $\mu_{xx}$ and $\mu_{xy}$. Experiments typically take place in the high magnetic field regime $\mu_{xx} \ll \mu_{xy}$ and $\mu_{xy} \simeq B^{-1}$ (this corresponds to $\omega_c$ much faster than the scattering rate). Hence it is reasonable to first find the frequency of the resonant plasmon modes in the limit $\mu_{xx} = 0$. The potential $V_{t}$ can be determined from the time dependent density $n_{t}$ by solving the electrostatic Poisson equation, for simplicity we will assume for now a local electrostatics approximation $n_{t} = \chi V_{t}$ where the compressibility $\chi = \frac{4 \epsilon_0}{h e}$ is obtained from a plane capacitor model. Simulations with an exact solution of electrostatic equations will be presented later. The static electron density in presence of a density gradient can be written as $n_a(r, \theta) = n_0(r) + n_{0c}(r) \cos \theta$ where $(r, \theta)$ are polar coordinates on the helium surface oriented along the gradient direction. Away from the edges of the electron gas, we can approximate $n_{0c}(r) = \lambda r$ and treat the gradient $\lambda$ as a small anisotropy parameter. It is thus natural to expand $n_t$ and $V_t$ in harmonics of the angle $\theta$: 
\begin{align}
n_{t} = n_{t0}(r) + n_{tc}(r) \cos \theta + n_{ts}(r) \sin \theta 
\label{eq:nt}
\end{align}
where we have kept the lowest harmonics. This  procedure is justified since the only anisotropy comes from the uniform density gradient which couples only nearby harmonics through the $\cos \theta$ term. 
Expanding to the lowest polar angle harmonics, we cast Eq.~(\ref{dtne}) into an effective Schr\"odinger equation which describes standing-modes of electron-density oscillations:\begin{align}
\frac{\chi^2 \omega^2}{\mu_{xy}^2} \psi = -\frac{\lambda^2}{2} \partial_r^2 \psi + \frac{3 \lambda^2}{8 r^2} \psi  +  \frac{ \left(\partial_r n_0 \right)^2}{r^2} \psi
\label{eqSch}
\end{align}
in this equation we introduced the effective wavefunction $\psi(r) = \sqrt{r} n_{ts}$ and $\omega$ is the frequency of the density oscillation, its time dependence obeys $\partial_t^2 \psi = -\omega^2 \psi$. This equation 
describes a radial wave which propagates at velocity $v = \mu_{xy} \lambda/\sqrt{2} \chi$. As in quantum mechanics, the shape of the wavefunction $\psi$ is 
controlled by the external potential. From Eq.~(\ref{eqSch}) we see that it contains a term describing repulsion at the origin and a confinement term proportional to the square of the gradient of the static isotropic density distribution $(\partial_r n_0)^2$. The obtained plasmon mode exists only due to the simultaneous 
presence of a magnetic field and of the anisotropic density gradient $\lambda$, we will thus call it a magneto-gradient plasmon (MGP).  
The frequency, $\omega_g$, of first MGP mode is given by the ground state of the Schr\"odinger equation Eq.~(\ref{eqSch}), it scales as $\omega_g \sim \mu_{xy} \lambda/(\chi R)$  where $R$ is the radius of the electron cloud. This frequency vanishes the limit $R \rightarrow \infty$, its behaviour is thus similar to EMP
which also do not have a gap and can have frequencies much below $\omega_c$. The MGP frequency drops to zero at $\lambda = 0$ and will be 
overdamped if the density gradients are not strong enough. Fortunately due to the high mobilities of the electrons on helium system this mode can be visible even 
for small density gradients. Since Eq.~(\ref{eqSch}) is a standing wave equation, in addition to the lowest frequency mode $\omega_g$ resonances are expected 
around its harmonics $\omega_n = n \omega_g$ ($n \ge 1$ an integer), these harmonics will however turn out to be overdamped in our experiments. 

 The previous calculation showed that a small density gradient can create delocalised bulk magneto-plasmon modes well below the cyclotron resonance frequency, 
which is usually believed to give the gap frequency for bulk magneto-plasmons. This calculation was performed using a local density approximation $n_t = \chi V_t$, and does not provide a complete description of the low frequency magneto-plasmon modes. Indeed Eq.~(\ref{eqSch}) does not predict any finite frequency modes in the limit $\lambda \rightarrow 0$ and the edge magneto-plasmon modes are thus missing. Hence a more realistic theory, reproducing the already known magnetoplasmons is thus needed. Such a theory has to go beyond the local density approximation and treat the long range Coulomb interactions in realistic way. This requires to fix the electrostatic environment of the electron gas and its properties. From here we will focus on a realistic model of our experimental setup with electrons on helium.

\begin{figure}
\includegraphics[clip=true,width=8cm]{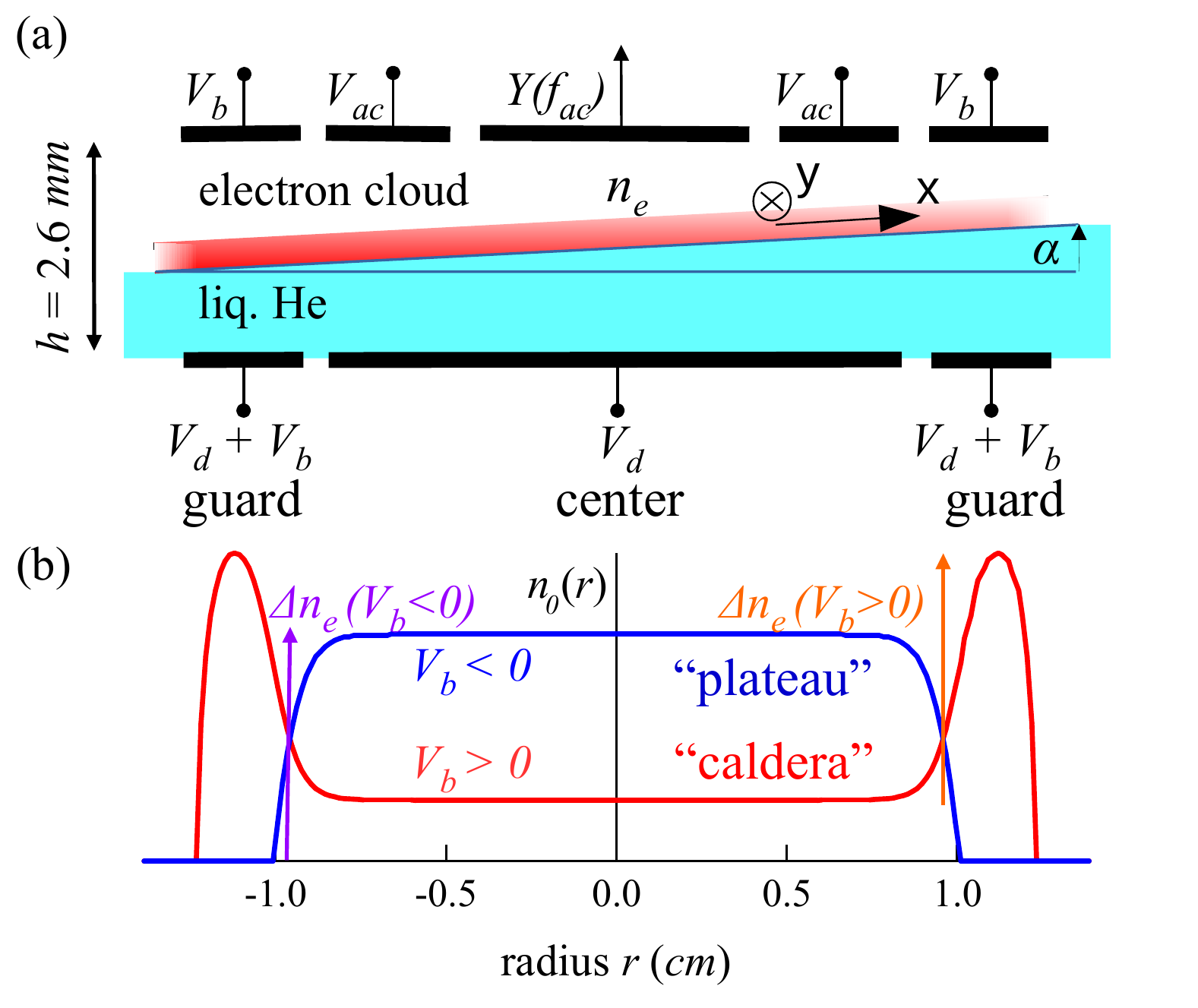} 
\caption{(a) Sketch of the experimental cell with applied DC and AC voltages. The helium cell and electrodes have cylindrical symmetry but the helium level is slightly tilted with an angle $\alpha$. A trapping potential $V_d = 7{\;\rm V}$ is applied to all bottom electrodes and a bias voltage $V_b$ can adjust the density between the outer guard and central regions. The cell admittance $Y(f_{ac})$ at frequency $f_{ac} = \omega/(2 \pi)$ is measured between top central and middle electrodes on which an AC potential $V_{ac}$ at frequency $f_{ac}$ is applied. Panel (b) shows the typical ``plateau'' and ``caldera'' profiles which occur respectively at $V_b < 0$ and $V_b > 0$. Orange and purple arrows illustrate the two possible definitions of $\Delta n_e$ in Eq.~(\ref{eqEMP}) for the two types of density profile.
}
\label{figGeometry}
\end{figure}

A sketch of the system is shown on Fig.~1, electrons are trapped on a helium surface by a pressing electric field. If the pressing electric field is perfectly perpendicular to the helium surface the geometry has cylindrical symmetry with respect to the polar angle $\theta$ and no gradient is present $\lambda = 0$. However a small misalignment angle $\alpha$ between the electric field direction and the normal of the helium surface leads to an in plane electric field component $\alpha E_\perp$ ($\alpha \ll 1$) which will create, within the local density approximation, a density gradient $\lambda = \chi \alpha E_\perp$. Since we assume that the helium surface remains flat in the region where electrons are confined, the drift diffusion equation Eq.~(\ref{dtne}) at the surface remain unchanged for finite $\alpha$ and it is only the relation between $n_t$ and $V_t$ which is changed when we use the exact non-local electrostatics. Since the normal of the electric field presents a discontinuity as it crosses the electrons on helium cloud, a direct perturbation theory expansion around the isotropic solution is not possible. We derived a suitable perturbation theory expansion by performing a transformation into a curved set of coordinates where the position of the interface remains fixed with $\alpha$. This expansion, to the lowest order, leads to a modified Laplace equation which is given in Appendix. Finite elements (FEM) \cite{FreeFem} simulations based on this equation, confirm the validity of the approximation $n_{0c}(r) \simeq \alpha \chi E_\perp r$ for all the shapes of the electron cloud explored experimentally except near the edge of the electron cloud. Thus a small inclination of the helium surface with respect to the helium cell creates a well defined density gradient which only weakly depends on the shape of the electron cloud $n_0(r)$. To simulate approximately the AC response of electrons on helium, we used the Poisson equation on AC potential in the limit $\alpha = 0$ and drift diffusion equations on the helium surface to determine the expected admittance of the cell. The full equations and a discussion on their formal validity are provided in Appendix.

\begin{figure}
\centering
 \includegraphics[clip=true,width=\columnwidth]{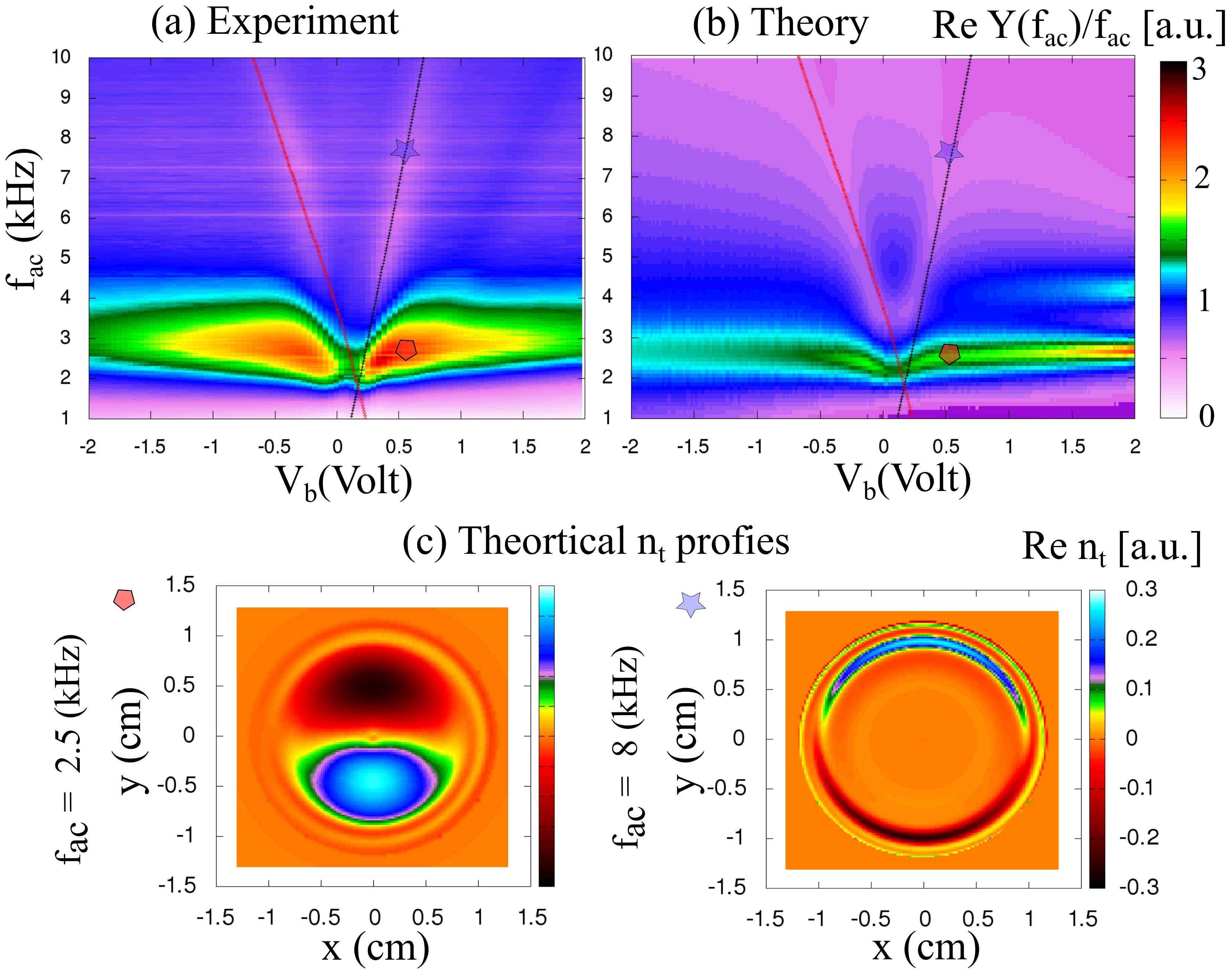}
 \caption{(a) Magnetoplasmon modes appear as peaks in the real part of the cell admittance $Y(f_{ac})$ which is shown here as function of the density profile (controlled by the bias $V_b$ between guard and central regions) and excitation frequency $f_{ac}$. Two magneto-plasmon modes are observed in the explored frequency range with very different dependence on $V_b$: a dispersing mode at higher frequency (branch carrying the star symbol) and a low frequency mode (branch carrying the polygon symbol) with very small dependence of the electron cloud density profile. (b) Finite element simulations of $Y(f_{ac})$ based on Eq.~\ref{dtne} taking into account a small tilt $\alpha = 0.4\;{\rm deg}$ (fitted to data). (c) Theoretical oscillating density profiles ($n_t$), the low frequency ``magneto-gradient'' plasmon ($n_t$ at the polygon symbol) is delocalised across all the electron cloud, whereas the higher frequency plasmon ($n_t$ at the star symbol) is an edge magneto-plasmon (or inter-edge at $V_b > 0$)  propagating in one-dimension, the two modes seem to make an avoided crossing when their frequencies overlap near $V_b = 0$.
 }
\label{figEMPandMGP}
\end{figure}

We now present experiments that reveal the co-existence of 1D EMP and 2D MGP plasmons, the simulations, which are presented simultaneously, will allow to confirm the identification of the observed modes. The experiments were realised on an electron cloud (see Fig.~1) with $N_e = 3\times 10^7$ electrons at a magnetic field of $B = 0.3\;{\rm Tesla}$ and temperature of $300$mK. The AC transport response of the electron cloud is measured in a Corbino geometry using a Sommer-Tanner method. 
An AC excitation voltage $V_{ac}$ with a $30$mV amplitude is applied on the intermediate top-ring electrode at a frequency $f_{ac}$ (1 to 10 kHz) and the induced pick-up signal from the top central electrode is then measured with a voltage amplifier and a lock-in detector giving the AC cell admittance $Y(f_{ac})$. The position of magneto-plasmon resonances then shows as peaks in the admittance $Y(f_{ac})$ for a fixed bias voltage $V_b$. This bias between the outer guard and central electrodes can tune the frequency of the magneto-plasmons by controlling the shape of the electron cloud \cite{Alexei2015,Alexei2019}. For $V_b < 0$ the electron cloud adopts a ``plateau'' density profile, where the density $n_0(r)$ is a monotonously decaying function of the radial distance to the cloud center $r$, while for $V_b > 0$ the electron density takes a ``caldera'' profile which has a density maximum inside the guard region at edge of the electron cloud (Fig.~1b). The EMP which couple to the top measuring electrodes are different in the two regimes. For a ``plateau'' profile the outer-edge of the cloud is closer to the center and it is the EMP propagating at the perimeter of the cloud which are detected. For a ``caldera'' geometry the cloud expands and the outer-perimeter becomes weakly coupled to the measurement electrodes, instead it is the inter-edge magneto-plasmon at the boundary between the guard and central regions which is more easily excited. To simplify further discussions we will describe both situations as an EMP mode. The theoretical expressions for the propagation velocity are indeed similar in both cases \cite{Volkov1988,Volkov1991}:
\begin{align}
v_{EMP} \simeq \frac{\Delta n_e}{2 \pi \epsilon_0 B} \ln \frac{1}{q h}
\label{eqEMP}
\end{align}
where $q$ is the wave vector ($q = 1/R$ for the lowest frequency mode where $R$ is the radius of central bottom electrode) and $\Delta n_e$ the difference in electron density between the center and guard regions (this definition is discussed in more precisely below). 

For the magneto-gradient plasmon the propagation velocity is given by:
\begin{align}
v_{MGP} \simeq \frac{\alpha E_\perp}{B \sqrt{2}}
\label{eqMGP}
\end{align}
this expression is obtained from Eq.~(\ref{eqSch}) using the approximation $\lambda = \alpha \chi E_\perp$. It depends on the perpendicular electric field $E_\perp$ but not on $\Delta n_e$ as opposed to the EMP modes. 
This difference in the $\Delta n_e$ dependence provides a convenient method to distinguish between MGP and EMP modes. We represent the admittance $Y(f_{ac})$ as function of both $V_b$ and $f_{ac}$ as a color scale map that allows to visualize the dependence of the mode frequency on the voltage $V_b$ (and thus on $\Delta n_e$). Modes that ``disperse'' as function of $V_b$ are candidate 1D EMP modes while the absence of a $V_b$ dependence suggests a magneto-gradient mode. 

Our experimental results are shown on Fig.~\ref{figEMPandMGP} together with FEM simulations of the perturbation theory that we introduced. Both experiment and simulations show the presence of two resonant modes at low frequency. A mode whose frequency that strongly depends on $V_b$ with a dependence that reminds the dispersion relation of ``Dirac'' fermions, and a second mode whose frequency is almost independent on $V_b$ except when its frequency crosses the frequency of the dispersive mode. 

To identify the dispersive mode as an EMP we show the theoretical frequency expected from Eq.~(\ref{eqEMP}). For the ``calderia'' 
geometry (positive $V_b$, black lines), we set $\Delta n_e$ as the difference between the electron density on top of the rim in the guard region 
and the density in the center of the electron-cloud, it reproduces the experimental EMP frequency without adjustable parameters. The EMP modes should not be visible in Corbino geometry, the FEM simulations show that a small tilt of the helium cell can explain their contribution to the Corbino signal with an amplitude comparable with the experiment. For the plateau geometry (red line for negative $V_b$) the density in the guard entering in $\Delta n_e$ had to be reduced by 20\% compared to its maximum density in the guard. This phenomenological correction probably reflects a more complex situation where the boundary of the electron cloud boundary moves with $V_b$ as the cloud is pushed towards the center. FEM simulations predict the correct position for the EMP mode without adjustable parameters even in this case. The predictions for the linewidth and admittance amplitude are less accurate as they dependent on the ratio $\mu_{xx}/\mu_{xy}$ which was assumed to be fixed to $5 \times 10^{-3}$ and without any density dependence. To confirm the 1D character of this dispersive mode we also represented the simulated oscillating density profile $n_t$ of the EMP mode on Fig.~\ref{figEMPandMGP}, the oscillating density is indeed localised in a narrow strip of width $h$ at the boundary between the central and guard regions. Note that, the perturbation theory is not reliable at the outer edge of the electron cloud and the peak in $n_t$ on the outer cloud boundary may not be physical.

The lowest frequency mode on Fig.~\ref{figEMPandMGP} has a resonance frequency which is independent of $V_b$ except near the crossing points with the previously identified EMP mode. It is thus a candidate magneto-gradient plasmon (MGP) mode. To confirm this assignment, we checked that this frequency scales as $\propto E_\perp/B$ as expected from Eq.~(\ref{eqMGP}). 
The only other parameter in Eq.~(\ref{eqMGP}) is the inclination of the helium free surface $\alpha$ compared to the electric field. It was not possible to control this angle precisely in our experiment however we confirmed that this frequency changes indeed with a small variation (of around 0.1 deg) of the fridge inclination. These experiments are shown in appendix (see Fig.~\ref{figAppendixTilt}). The FEM simulations allow to visualize the oscillating density profile which is displayed on Fig.~\ref{figEMPandMGP}, this mode is delocalised across the entire electron cloud. In the appendix Fig.~\ref{figAppendixSegs}, we present experiments with segmented pick-up electrodes which allow to confirm some features of the MGP density distribution. A good agreement between simulations and the experimental GMP mode is obtained for $\alpha = 0.4$deg. At the crossing between MGP and EMP at $V_b = 0$ both simulations and experiment suggest an avoided crossing which implies the exciting possibility of realizing hybrid states with simultaneous characteristics of a one dimensional topological EMP and a delocalised two dimensional mode (note that changes in the GMP confinement potential also contribute to the change of the GMP frequency at $V_b=0$, this contribution is discussed in appendix Fig.~\ref{figTheoSchExpTheo}. Concerning the linewidth of MGP modes the simulations predict a similar linewidth to the edge magnetoplasmon modes. This contrasts with the experiment where the MGP is significantly broader than the EMP with a linewidth that depends on $V_g$. This could be due to the dependence of the mobility on the electron density which is not taken into account in the model. Indeed MGP is delocalized and thus can be more sensible to mobility gradients which appear as a result of the density variations across the electron cloud. In the data in Fig.~\ref{figEMPandMGP} the quality factor of the GMP resonance is not so high, however in appendix we shows experiments the quality factor increases substantially with stronger perpendicular field as shown in Fig.~\ref{figAppendixTilt}, tilt angle(Fig.~\ref{figAppendixEperp}) or at minima of microwave induced resistance oscillations (Fig.~\ref{figAppendixRF1}), proving that GMP is a genuine resonance of the 2D electron system. 

To summarize, we have shown  both theoretically and experimentally that a
system of electrons on the surface of liquid helium hosts a novel type
of bulk  excitation, the delocalized  two--dimensional magnetogradient
plasmon  modes  which  appear  in  the presence  of  a  small  density
gradient. This  system provides  a highly controllable  environment in
which the  interaction of  this novel  excitation with  the previously
known  (topological)  one--dimensional  edge  magnetoplasmons  may  be
studied. As gradients of carrier concentration can easily be present in mesoscopic devices, this 
mode can also be present in new two dimensional electrons systems . 

We acknowledge fruitful discussions with D.L. Shepelyansky and M. Dykman and thank ANR SPINEX for financial support.

\bibliographystyle{apsrev4-1}
\bibliography{hebib}

\section*{Appendix}
\setcounter{equation}{0}
\renewcommand{\theequation}{A\arabic{equation}}
\setcounter{figure}{0}
\renewcommand\thefigure{A\arabic{figure}}
\renewcommand{\figurename}{Appendix Figure}

\section*{A1. Derivation of the main equations}
\subsection{Derivation of the effective Schr\"odinger equation} 

This equation is derived from the drift diffusion equations written in the local density approximation: 
\begin{align}
\partial_t \left\{\begin{array}{c}
                    n_{t0} \\
                    n_{tc} \\
                    n_{ts}
\end{array} \right. =
\frac{\mu_{xy}}{\chi r} \left\{\begin{array}{c}
           \lambda \partial_r( r n_{ts} ) / 2\\
          n_{ts} \partial_r  n_0 \\
           \lambda r \partial_r n_{t0}- n_{tc} \partial_r n_0 
           \end{array} \right.
\label{dtnecs}
\end{align}
The density gradient $\lambda$ introduces a coupling between different angular harmonics of the electron density, introducing the effective wavefunction $\psi(r) = \sqrt{r} n_{ts}$ this equation can be reduced to the effective Schr\"odinger equation: 
\begin{align}
\frac{\chi^2 \omega^2}{\mu_{xy}^2} \psi = -\frac{\lambda^2}{2} \partial_r^2 \psi + \frac{3 \lambda^2}{8 r^2} \psi  +  \frac{ \left(\partial_r n_0 \right)^2}{r^2} \psi
\end{align}
where $\lambda = \frac{4 \epsilon_0 \alpha E_\perp}{h e}$ which is Eq.~3 from the main text. We remind that the only unknown parameter in this theory is the cell tilt angle $\alpha$,  we set it to $\alpha = 0.4$ deg as in the finite elements simulations. 

\begin{figure}[!htb]
\centering 
 \includegraphics[width=\columnwidth]{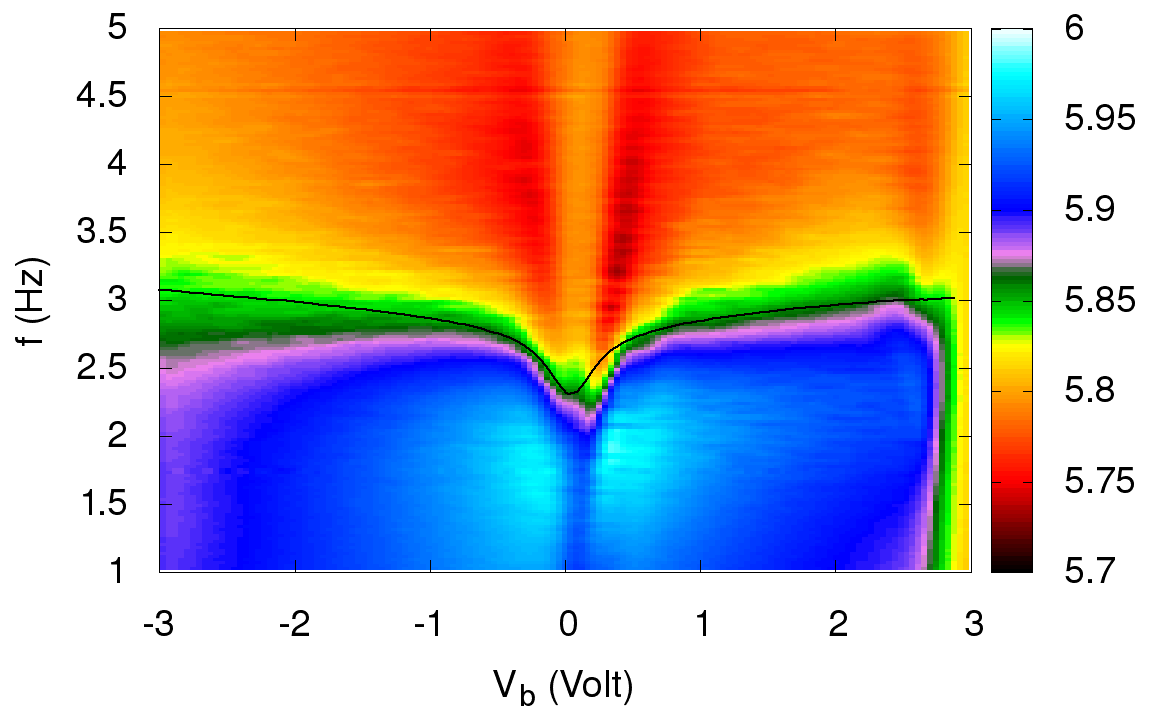}
\caption{\label{figTheoSchExpTheo} Imaginary part of the cell admittance ${\rm Im Y}/f_{ac}$ (in arbitrary units) for the experimental parameters of Fig.~2, the admittance of the cell mainly behaves as a capacitance and is thus divided by $f_{ac}$ to compensate its linear increase with frequency. The position of the MGP resonance is given by the green-contour curves, the lowest eigen-frequency of the effective Schr\"odinger equation for $\alpha = 0.4$ deg is shown by the black line. }
 \end{figure}

Fig.~\ref{figTheoSchExpTheo} shows a comparison between the experimental position of the MGP resonance and the predictions of the effective Schr\"odinger equation for the data from Fig.~2 (from the main text). The position of the resonance is more easily followed on the out of phase response, as it corresponds to a constant value contour. The color on Fig.\ref{figTheoSchExpTheo} thus shows ${\rm Im}\ Y$ (instead of ${\rm Re} Y$ on Fig.~2 as function of the bias voltage $V_b$ and excitation the position of the resonance. The position of the resonance corresponds to the green contour curves. We see that Eq.~3 gives a satisfactory description of the dependence of the resonance frequency as function of $V_b$. However the effective theory does not capture the double resonance structure near $V_b=0$ since it appears due to the crossing between the MGP and EMP modes which is not taken into account in this simplified model.

\subsection{Effective Poisson equation in deformed coordinates where the position of the helium layer is fixed.}

\begin{figure}[!htb]
\centering
\begin{tabular}{c}
 \includegraphics[width=\columnwidth]{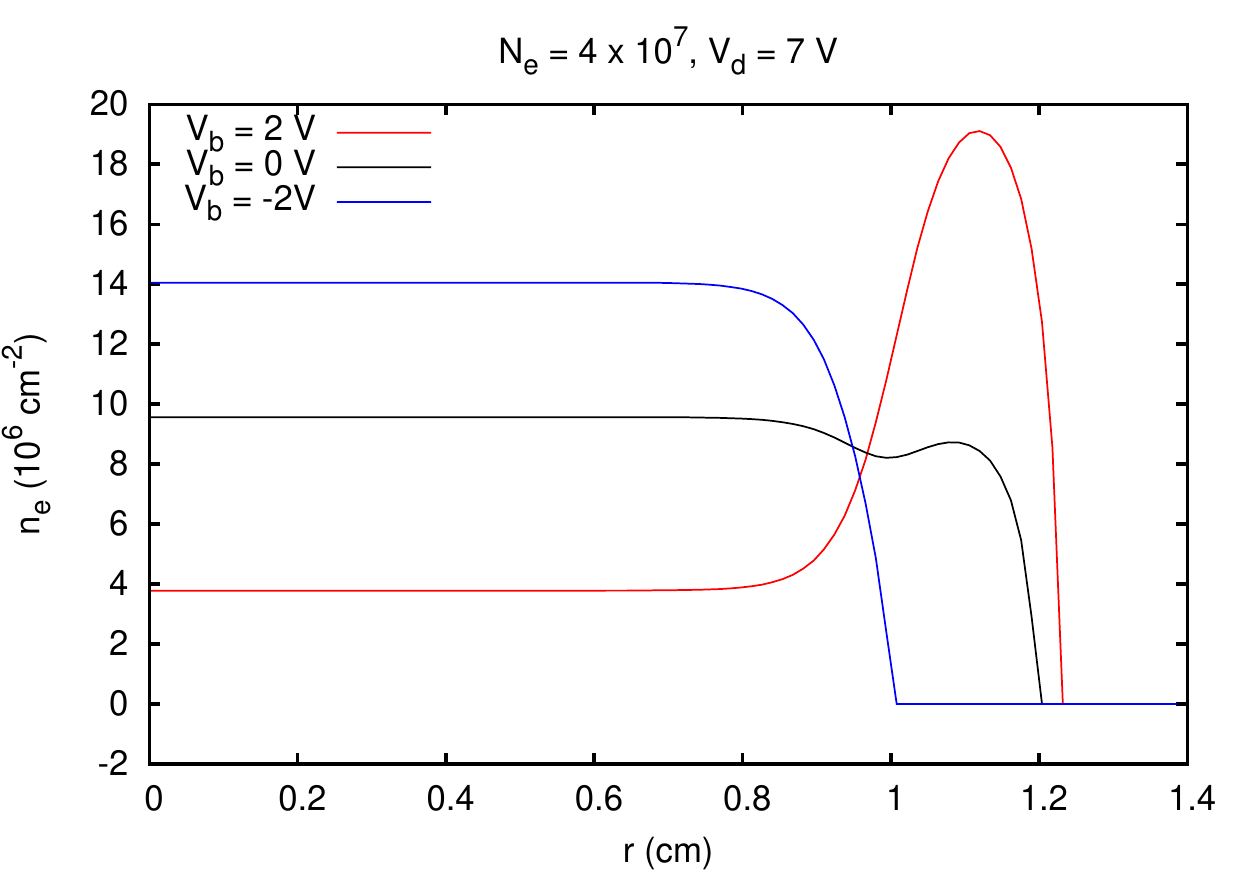} \\
 \includegraphics[width=\columnwidth]{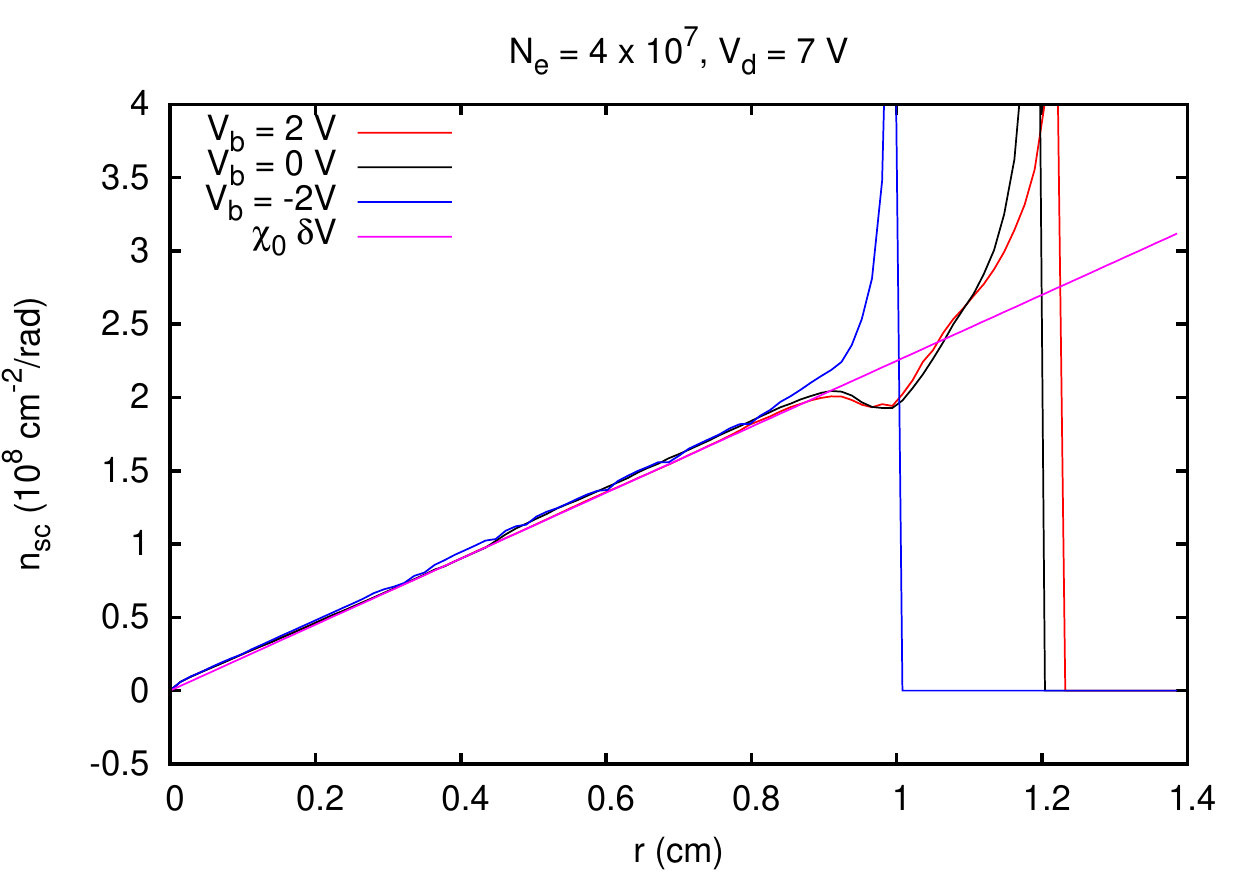}
\end{tabular}
 \caption{\label{figAppendix1} The top panel shows the evolution of the radial steady state density $n_0(r)$ for different bias voltages $V_b$ between central an guard reservoirs. The bottom panel shows the tilted component of density $n_{0c}$ under the same conditions (we remind that the total steady state density is the sum 
$n_0(r) + \alpha n_{0c}(r) \cos \theta$ ), $n_{0c}$ depends only weakly on $V_b$ in contrast to $n_0(r)$ and is well approximated by $n_{oc} = \chi \alpha E_\perp r$ (straight line).
 }
\end{figure}

To enable the use of standard perturbation series we thus need to perform a coordinate transformation which levels the 
helium surface in the cell while keeping in place the top and bottom electrodes. We choose a rotation in the $(x,z)$ plane (containing the electric field 
direction $z$ and the slightly miss-aligned helium surface normal) with a height dependent rotation 
angle $\phi(z) = \alpha \left(1 - \frac{4 z'^2}{h^2} \right)$. The transformation between coordinates is the realized by:
\begin{align}
 \left(\begin{array}{c}
x \\
z 
\end{array}\right) = 
 \left(\begin{array}{c}
\cos \phi(z') x' - \sin \phi(z') z' \\
\sin \phi(z') x' + \cos \phi(z') z'
\end{array}\right)
\end{align}
where $x'$, $z'$ are the new coordinates.

In the new coordinates the Poisson equation, to first order in $\alpha$, becomes:
\begin{align}
&\Delta V + \frac{8 \alpha}{h^2} \left( x \partial_z V + 2 x z \partial_{z}^2 V - 3 z \partial_x V - 2 z^2 \partial_{xz} V  \right) = 0 
\label{EqPoissonTilt}
\end{align}
in this form it can be expanded in powers of $\alpha$. We see from Eq.~(\ref{EqPoissonTilt}) that $\alpha$ induces a coupling only between neighboring angular harmonics,
 thus to first order in $\alpha$ to which we will limit ourselves here, only $\cos \theta$ and $\sin \theta$ terms will be generated.

We solved Eq.~(\ref{EqPoissonTilt}) for a stationary electron cloud without AC excitation, this allows us to find the static density profile 
$n_a(r, \theta) = n_0(r) + \alpha n_{0c}(r) \cos \theta$ induced by the tilt of the cell. In the stationary case the potential of the electron cloud is 
constant and fixed by the total charge of the cloud. In the isotropic case the problem then reduces to find a stable boundary of the electron 
could for which the electric field at the boundary vanishes, which was done in a systematic way for different geometries in \cite{Alexei2015}. 
To find the anisotropic correction $n_{0c}(r)$ we iterated Eq.~(\ref{EqPoissonTilt}) neglecting the small deformation of the circular cloud boundary. 
While this approach should give accurate predictions in the center of the electron cloud, the validity of the perturbation theory breaks down near the cloud boundary. 
The results of our finite element calculations for $n_{0}(r)$ and $n_{0c}(r)$ are shown below on Fig.\ref{figAppendix1}.

To be fully consistent we in principle need to use the modified Poisson equation Eq.~(\ref{EqPoissonTilt}) which introduces a mixing between harmonics due to the tilt, 
we found however that using the usual Poisson equation $\Delta V = 0$ gives already a good description of the experiment. 
This probably comes from the fact that the dynamic drift-diffusion equations already introduces a coupling between modes through the density gradient $n_{0c}$ and many 
(but seemingly not all) of the terms that would come from Eq.~(\ref{EqPoissonTilt}) become second order in $\alpha$.

\subsection{Full drift diffusion equations for the lowest angular harmonics} 

For reference we write the full drift diffusion equations as a function of the steady state radial distribution $n_0$ and the density gradient $n_{0c}$ including contributions from both $\mu_{xx}$ and $\mu_{xy}$, these are the equations which are solved to build Fig.~2 in the main text.

The AC potential $V_{t}$ is decomposed into its lowest harmonics $V_{t} = V_{t0}(r) + V_{tc}(r) \cos \theta + V_{ts}(r) \sin \theta$. 

\begin{widetext}

\begin{align}
  \partial_t
  \left\{
  \begin{array}{c}
    n_{t0} \\
    n_{tc} \\
    n_{ts}
  \end{array}
  \right\}
  =
  \frac{\mu_{xy}}{r}
  \left\{
  \begin{array}{c}
    \frac{1}{2} \partial_r( n_{sc} V_{ts} )  \\
    V_{ts} \partial_r  n_0 \\
    n_{sc} \partial_r V_{t0} - V_{tc} \partial_r n_0 
  \end{array}
  \right\}
  +\frac{\mu_{xx}}{r}
  \left\{
  \begin{array}{c}
    \partial_r
    \left[
       r(n_0\partial_rV_{t0}+\frac{1}{2}n_{sc}\partial_rV_{tc})
    \right]
    \\
    \partial_r
    \left[
       r(n_0\partial_rV_{tc}+n_{sc}\partial_rV_{t0})
    \right]
    -\frac{n_0}{r}V_{tc}
    \\
    \partial_r
    \left[
       r(n_0\partial_r V_{ts})
    \right]
    -\frac{n_0}{r}V_{ts}
  \end{array}
  \right\}
\label{FullFEM}
\end{align}

\end{widetext}

\section*{A2 - probing the spatial structure of the MGP plasmon}

\begin{figure}[!htb]
\centering 
\begin{tabular}{c|c}
${\rm Re}\;n_t$ & ${\rm Im}\;n_t$ \\
 \includegraphics[trim=130 100 150 100,clip,width=0.5\columnwidth]{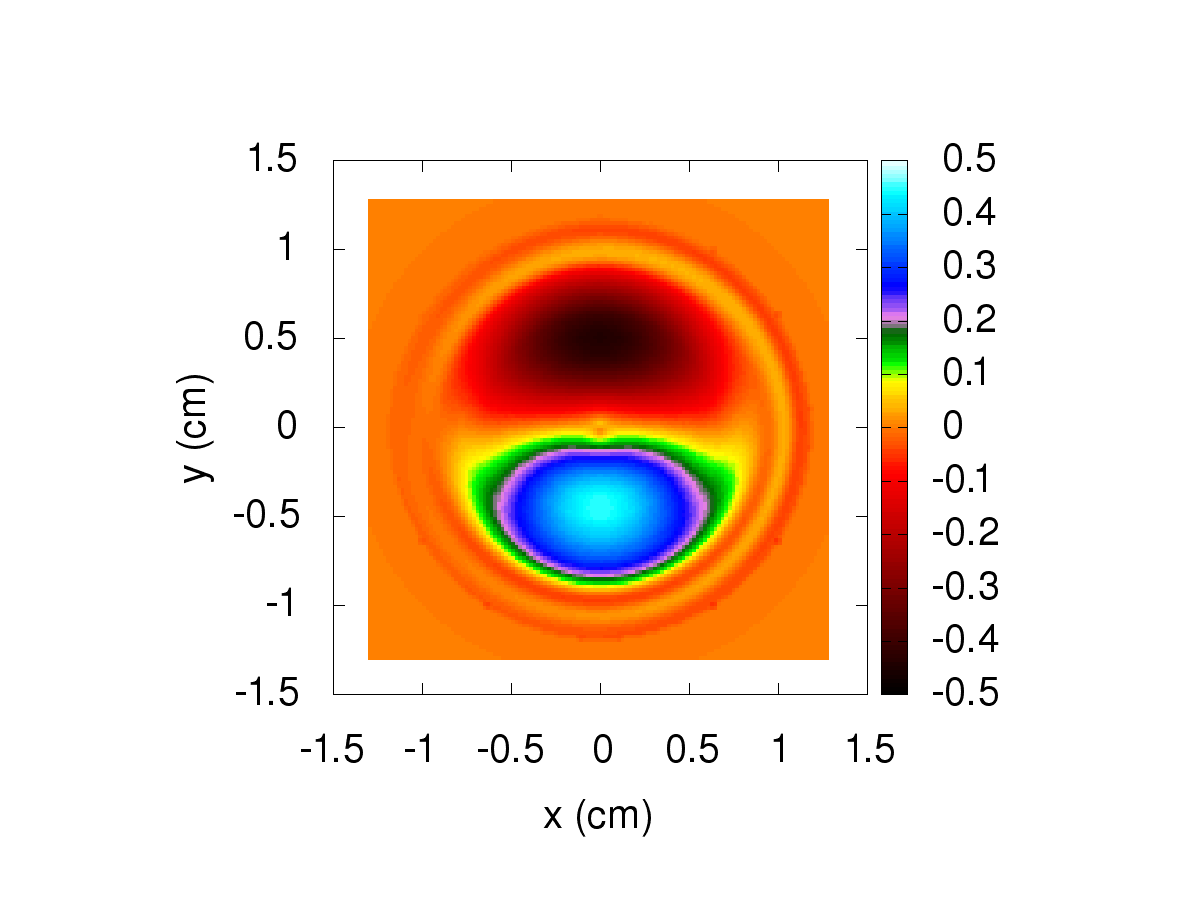} & \includegraphics[trim=130 100 150 100,clip,width=0.5\columnwidth]{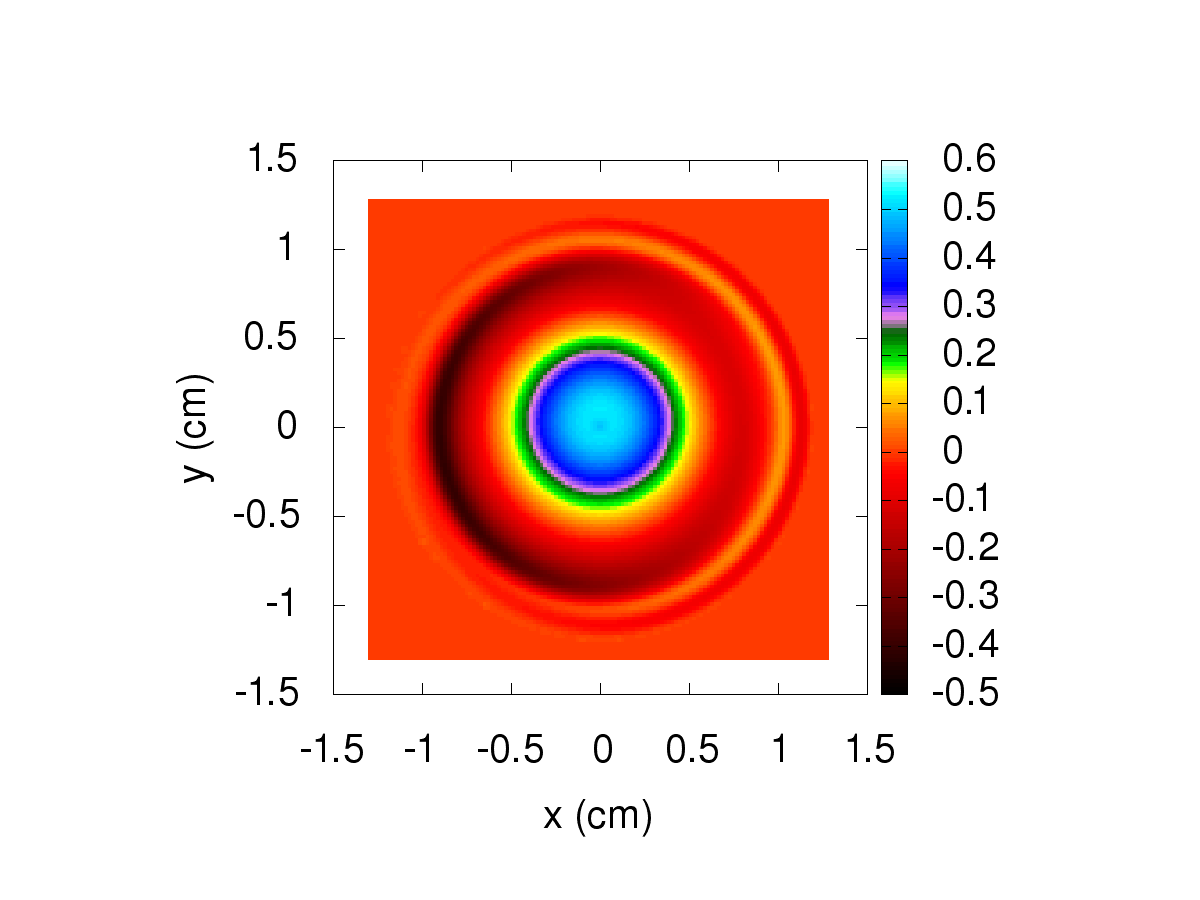} 
\end{tabular}
\caption{\label{figAppendixNt} Time-dependent  part of  the electron  density as  a function  of the spatial coordinates x and y,  calculated for the following parameters:  $V_{sat} = 0.8$ Volt, $V_{b} = 0$ Volt, $V_d = 7.5$ Volt, $B = 0.23$Tesla and $f_{ac} = 2.5$ kHz (pentagon symbol in Fig.~2 from the main text). The electron-density has both an in-phase component ${\rm Re} n_t$ (left
panel) and an out-of-phase component ${\rm Im} n_t$ (right panel) with respect to the excitation voltage. }
\end{figure}

The system of Eqs.~(\ref{dtnecs}) gives a relation between the time derivative of the polar angle average of the electron density $\partial_t n_{t0}$ and the radial derivatives of its first angular harmonic $n_{ts}$:
\begin{align}
\partial_t n_{t0} &= \frac{\lambda \mu_{xy}}{2 \chi r} \partial_r( r n_{ts} ) 
\label{sintoiso}
\end{align}
This relation gives an insight on the spatial structure of the MGP mode which is most easily visualized by plotting both the in-phase and out-of phase density maps which are both displayed on Fig.~\ref{figAppendixNt}, note that Fig.~2 from the main text for compactness only showed ${\rm Re}\;n_t$. We see that while the in phase-density response is dominated by the $\sin \theta$ angular harmonic, the out-of phase component is essentially isotropic.

\begin{figure}[!htb]
\centering 
\begin{tabular}{c|c} 
\multicolumn{2}{c}{  \includegraphics[width=0.4\columnwidth]{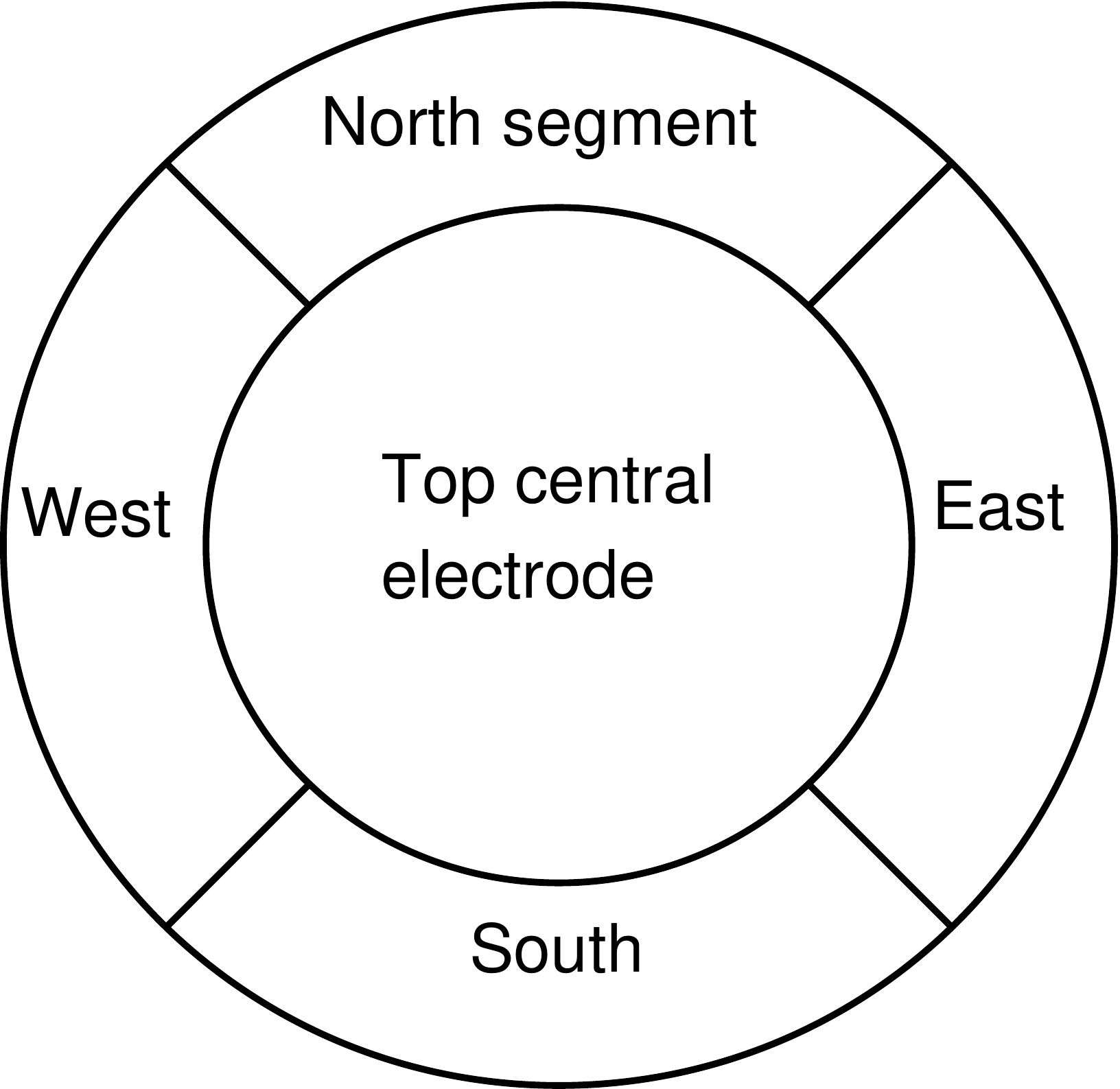}  }\\
Experiment & Finite elements simulations \\
  \includegraphics[width=0.5\columnwidth]{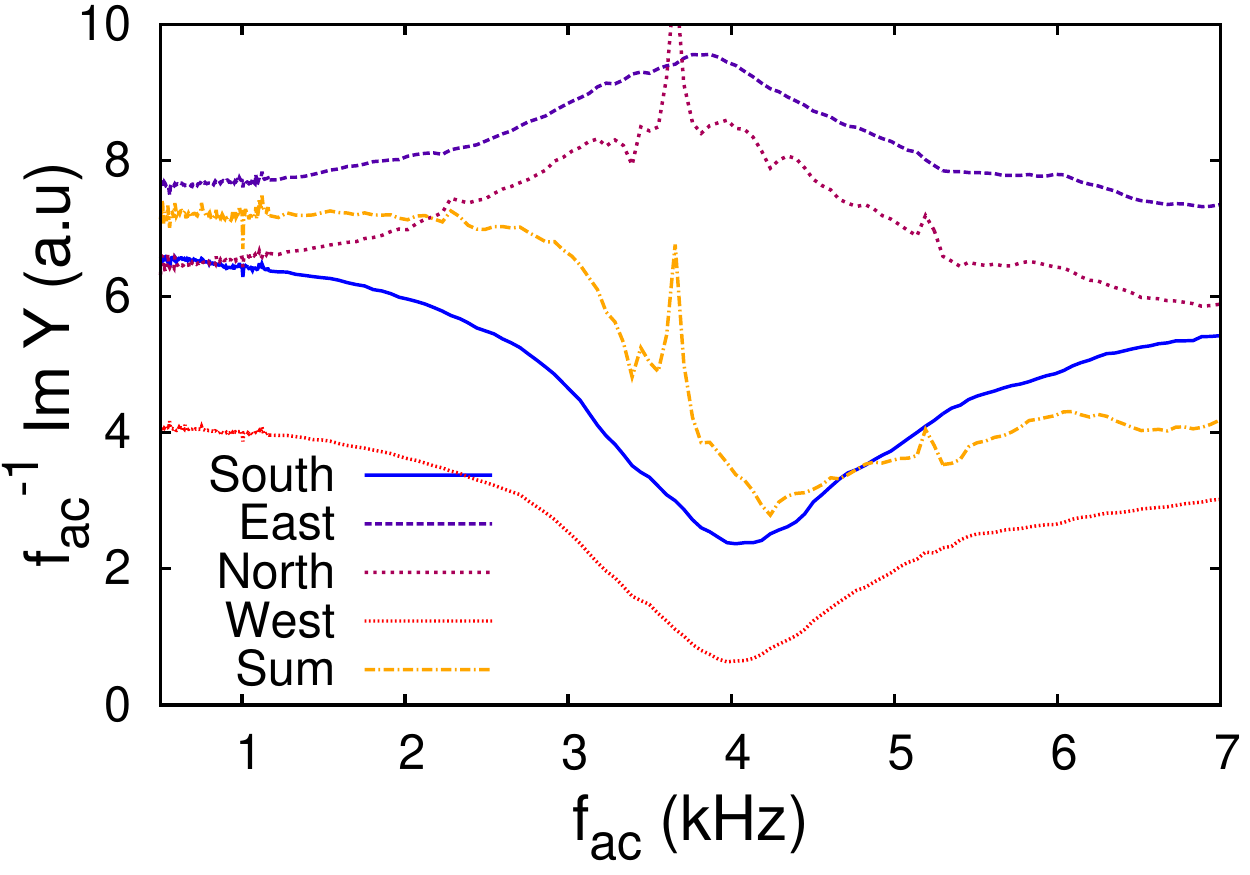} & \includegraphics[width=0.5\columnwidth]{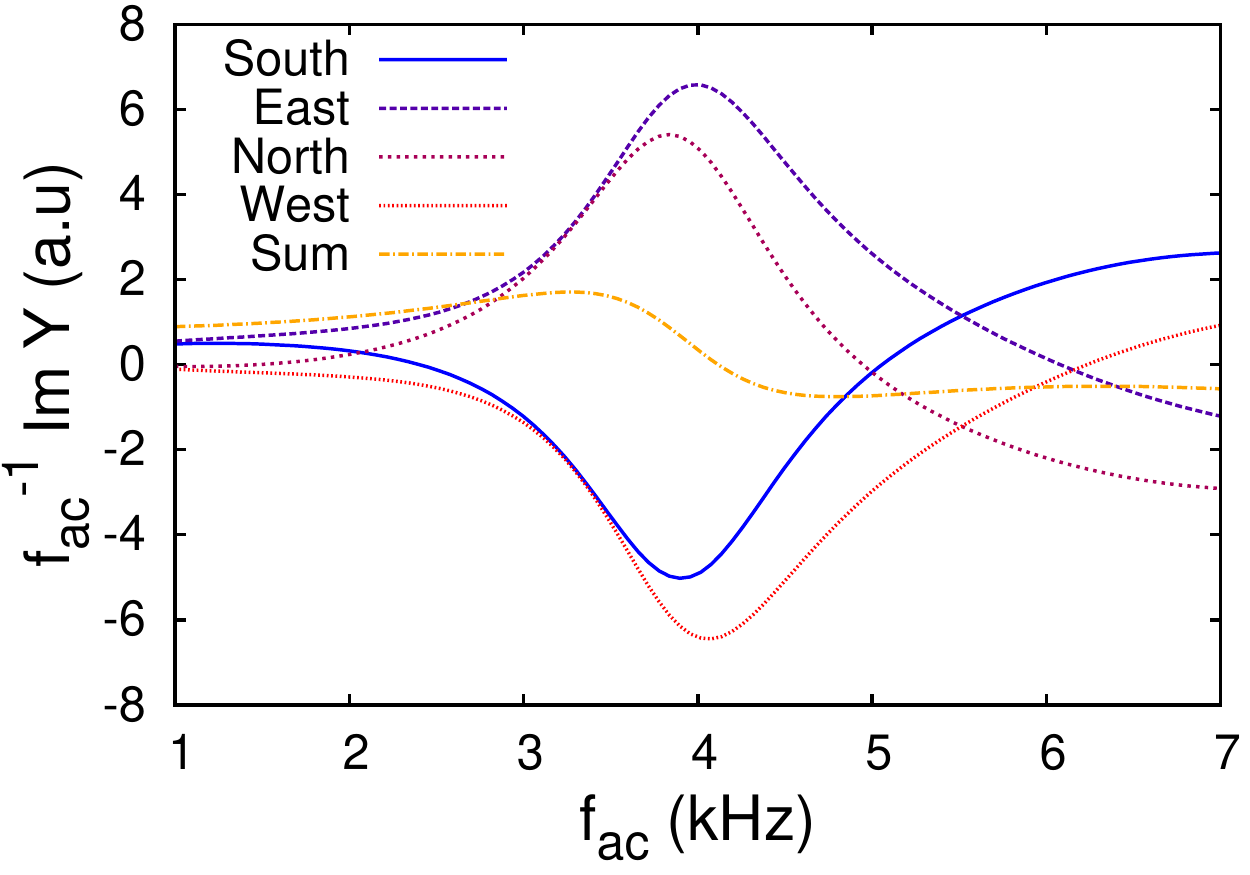} \\
  \includegraphics[width=0.5\columnwidth]{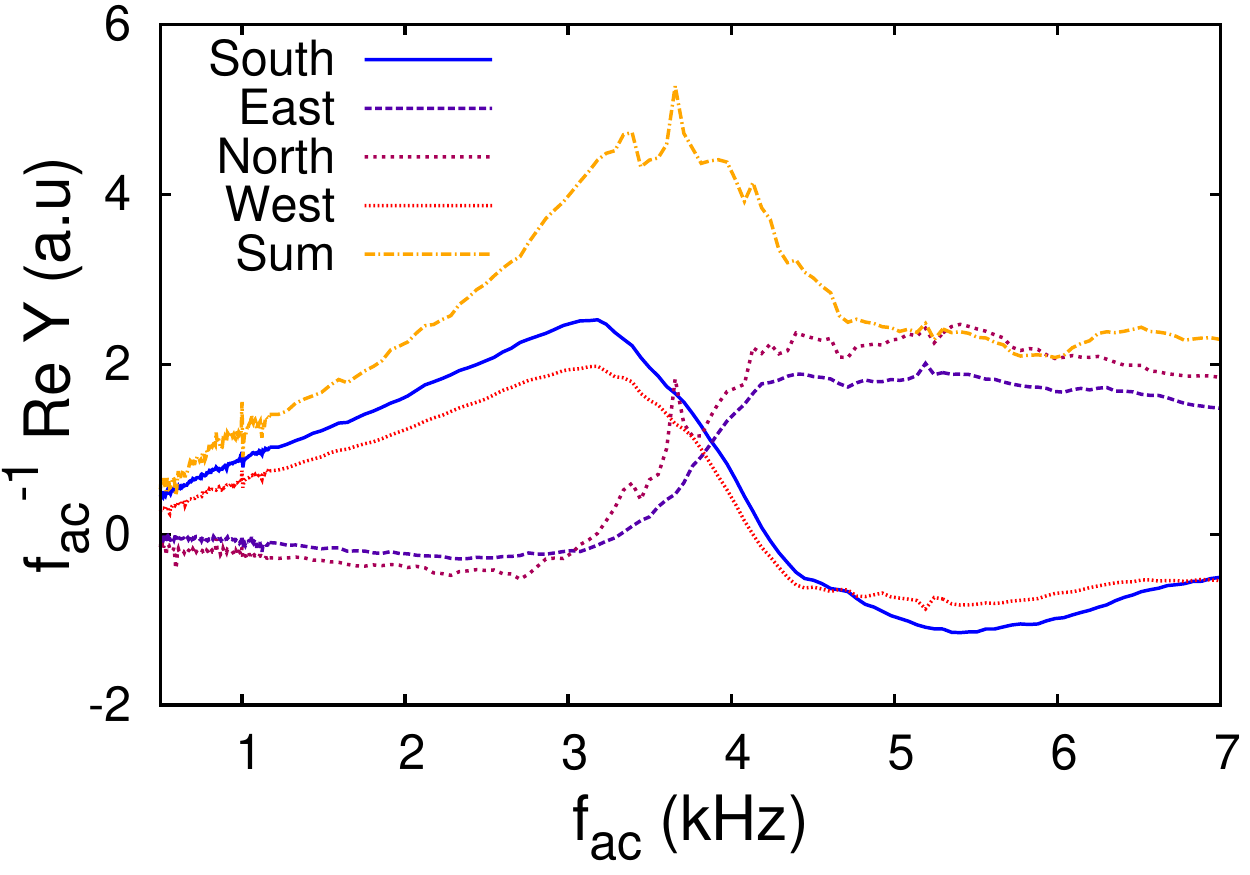} & \includegraphics[width=0.5\columnwidth]{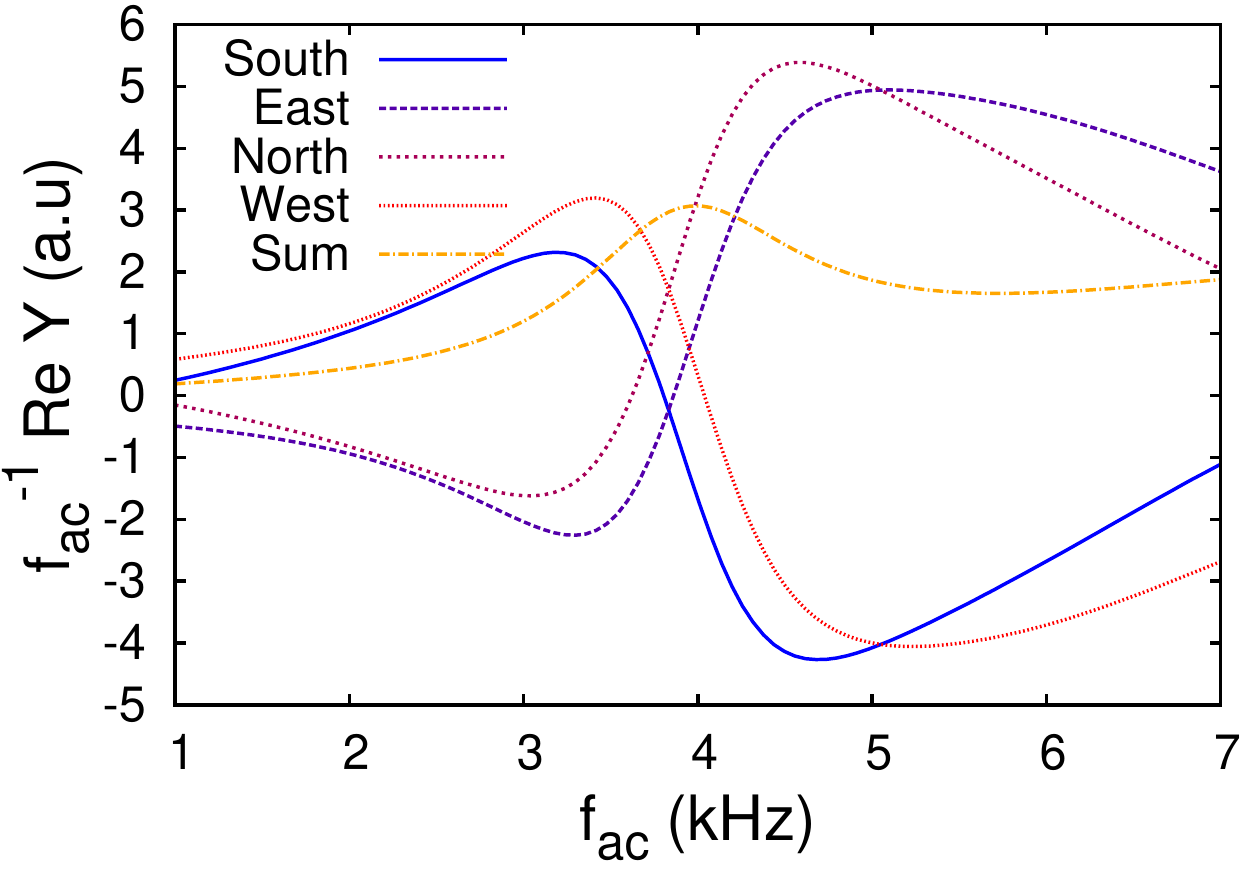}
\end{tabular}
\caption{\label{figAppendixSegs} Admittance measurements from a segmented electrodes, gate voltages were set to $V_{b} = 0$ Volt, $V_d = 7.5$ Volt, perpendicular magnetic field was $B = 0.23$Tesla and the total electron number was $N_e = 2 \times 10^7$, temperature was $0.2$ K. In the finite elements simulations we used the numerical values $\alpha = 0.5$ deg and $\mu_{xx}/\mu_{xy} = 2\times 10^{-2}$. The signal from individual segments and the isotropic signal (sum of all segments) are dephased by $90$ deg with each other, displaying different lineshapes peaked or dispersive respectively for the real/imaginary part of the admittance. This unusual behavior is reproduced by the finite element simulations, and illustrate the spatial structure of the oscillating electron density shown in Fig.\ref{figAppendixNt}. The signal is shown in arbitrary units since when measurements are done with a voltage amplifier the measured voltage depends on the capacitance of the measuring cables and a careful calibration is needed to recover the absolute value of the admittance. We have shown in \cite{Alexei2015} that with such calibrations the contribution of the electron gas to the measured capacitance perfectly agrees with the finite-elements simulations (except for zero resistance states). We thus feel justified here not to focus on the relative scales between the experimental and theoretical signals as we know that their amplitudes coincide in the limit of low frequencies and use arbitrary units for simplicity.  
}
\end{figure}

\begin{figure}[h]
\centering 
\includegraphics[width=0.9\columnwidth]{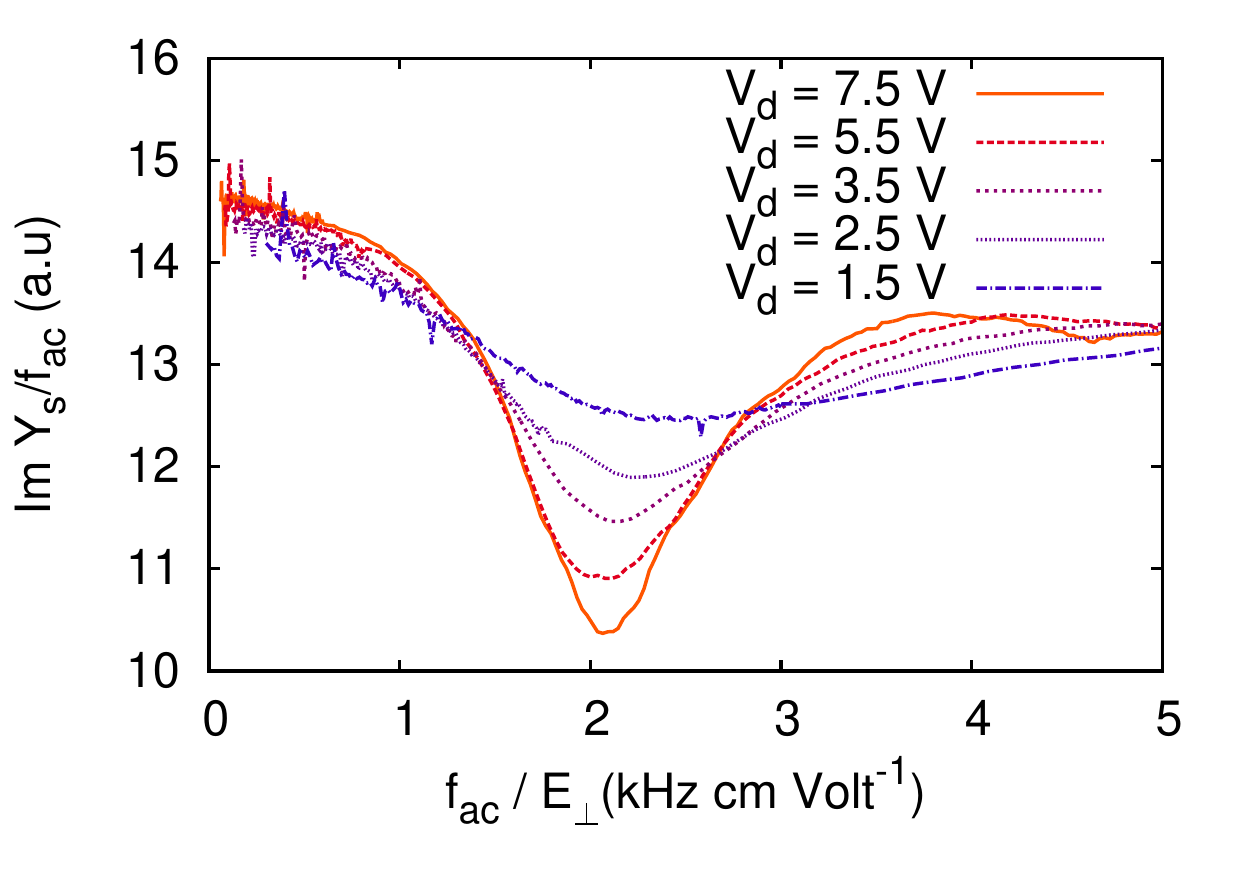}
\caption{\label{figAppendixEperp} This figure shows that the frequency of the GMP mode is proportional to the perpendicular electric field as predicted. The imaginary part of the admittance of one of the segments  (see discussion on the phase of the resonance in the previous section) is shown as function of the excitation frequency $f_{ac}$, the bias voltage is set to zero $V_{b} = 0$, and the perpendicular electric field is given by $E_\perp = V_d / h$ ($h$ = 2.6mm), $B = 0.23$Tesla. The excitation frequency $f_{ac}$ is divided by the perpendicular electric field, so that the position of the resonance remains fixed when $E_\perp$ is changed (different curves correspond to different $E_\perp$). The quality factor of the resonance increases with increasing $E_\perp$. }
\end{figure}

This observation can be checked in experiments with segmented Corbino electrodes which can probe not only the polar angle averaged electron density but also its angular harmonics. We performed such experiments using the electrode layout shown in Fig.~\ref{figAppendixSegs} where the exciting electrode on Fig.~1 (in the main text) was split into four segments (named after the cardinal directions). Compared to the experiments in the Corbino geometry the excitation was applied on the top central electrodes and four voltage amplifiers were used to collect the pick-up voltage from the segments measuring the admittance $Y_s$ between the central electrodes and the segments, as previously $Y$ is shown re-scaled by $f_{ac}$ due to the mainly capacitive behavior of the admittance. 

The results of the experiments are shown on Fig.~\ref{figAppendixSegs} together with finite elements simulations of the expected voltage on the segmented electrodes (obtained by angular integration of $n_{t0}, n_{tc}, n_{ts}$). We see that the spatial structure of the mode MGP gives rise to a rather counter intuitive behavior, the signal measured on individual segments seems out-of-phase with respect to the sum signal from all the segments. For example the imaginary part of the admittance from the segments is mainly an in-phase resonance line shape, with a peak at resonance, however the sum from all the segments behaves more like a derivative feature (dispersive out of phase signal). Similarly the real part of the admittance from individual segments behaves as a derivative (dispersive line shape) while the sum from all segments follows the line shape of an in phase resonance. This phase 90 degree phase shift between the admittance of a single segment and the isotropic average overall signal is nicely reproduced in the finite elements simulations and is a manifestation of the spatial structure of the oscillating charge density shown in Fig.\ref{figAppendixNt} where the $\sin \theta$ angular harmonic and the isotropic component are out of phase with each-other. 

\section*{A3 - perpendicular electric field dependence}

\section*{A4 - tilting the refrigerator}

 In the experiment shown on Fig.~\ref{figAppendixTilt}, we check that the frequency of the GMP mode can be changed by tilting the fridge which modifies the angle $\alpha$, the change of the resonance frequency is shown in the top panel with an increase of the resonance quality factor. The increase of the tilt angle was estimated from the shift of GMP mode. It was not possible to confirm this value independently since the design of our fridge was not optimized to allow the fine-tuning of its inclination with respect to the vertical direction (the estimated change is consistent with the indications of the bubble level meters that were fixed on the fridge). 
We are however certain that all other parameters remained fixed while the fridge was tilted. The gate voltages were fixed $V_{b} = 2$ Volt (in the top panel), $V_d = 7$ Volt and the trapped magnetic field remained $B = 0.3$ Tesla. The only parameter that could change is in fact the number of electrons trapped in the cloud as electrons could escape during the mechanical motion of the fridge. We confirmed that this was not the case in the bottom panel which shows scans of the cell admittance as function of the bias voltage $V_b$ before and after the tilt at $f_{ac} = 1.137$ kHz. The voltage threshold indicated by the black arrow at which the central region of the cell becomes filled with electrons is directly related to the total number of trapped electrons, it did not change before and after the tilt allowing us to confirm that $N_e = 3 \times 10^7$ did not change during the tilt of the fridge.

\section*{A5 - gradient magnetoplasmons under microwave irradiation}

We have shown that increasing the perpendicular electric field $E_\perp$ and the tilt angle $\alpha$ increase the quality factor of the GMP resonance, this probably occurs because the resonance frequency increases while the loss rate remains fixed. It is also possible to reduce the damping by applying microwave irradiation at a frequency which can induce zero resistance states. The modulation of the GMP resonance in this case is shown in Fig.~\ref{figAppendixRF1} for two values of $N_e$. When a microwave frequency at resonance between the lowest and first excited Rydberg state is sent into the cell (at the intersubband resonance in quantum well terminology) a $1/B$ periodic modulation of the quality factor of the GMP resonance is observed. At conditions corresponding to minima of the microwave induced resistance oscillations the quality factor of the GMP resonance is substantially increased. Fig.~\ref{figAppendixRF1} also shows the admittance $Y(f_{ac})$ without microwaves confirming that the frequency of the GMP resonance scales as $1/B$ as function of $B$, interestingly without microwave irradiation the quality factor of the resonance does not seem to depend much on the magnetic field. This is probably due to an approximate compensation between the $1/B$ decrease of the resonance frequency and the $1/B$ decrease of the ratio $\mu_{xx}/\mu_{xy}$. Experimentally this leads to a plateau where the admittance of the cell becomes independent on the magnetic field, thus due to the contribution of the GMP mode an admittance plateau can appear even when $\mu_{xx}$ still decreases with the magnetic field.

\begin{figure}[!htb]
\centering 
\begin{tabular}{c}
\includegraphics[width=0.9\columnwidth]{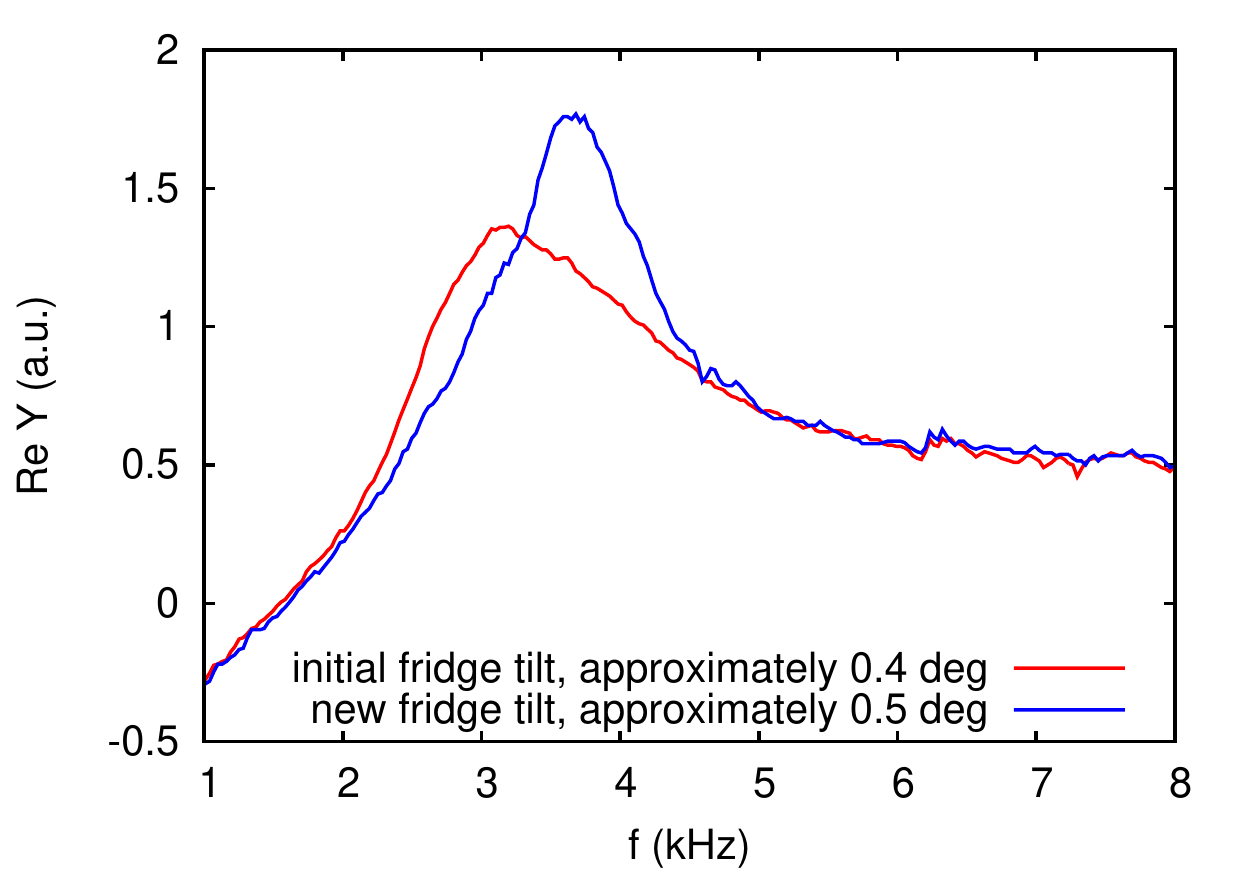} \\
\includegraphics[width=0.9\columnwidth]{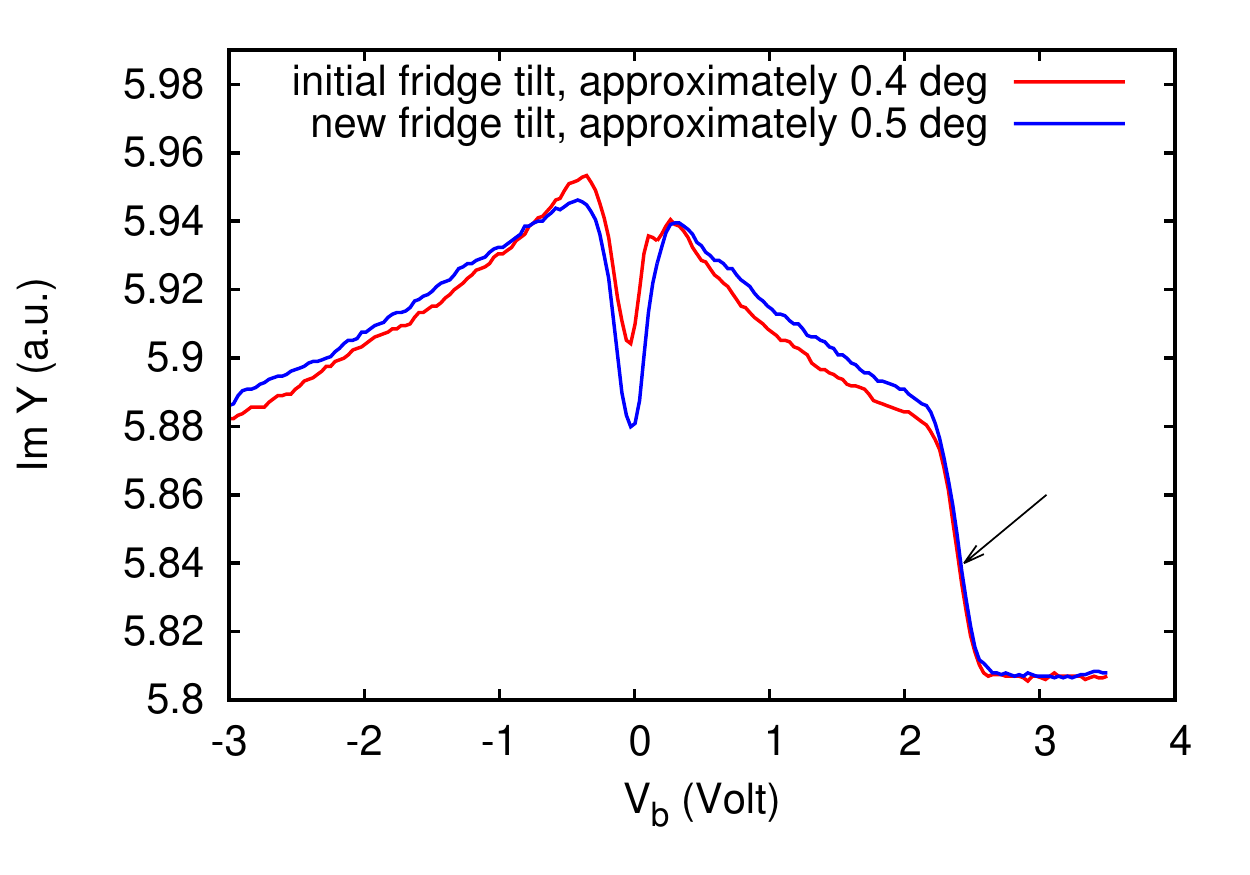}
\end{tabular}
\caption{\label{figAppendixTilt} The top panel shows that the frequency of the GMP mode can be changed by tilting the fridge which modifies the angle $\alpha$, the change of the resonance frequency is shown in the top panel with an increase of the resonance quality factor. The bottom panel confirms that the number of traped electrons did not change as the fridge was tilted.
}
\end{figure}

\begin{figure*}[!htb]
\centering
\includegraphics[width=2\columnwidth]{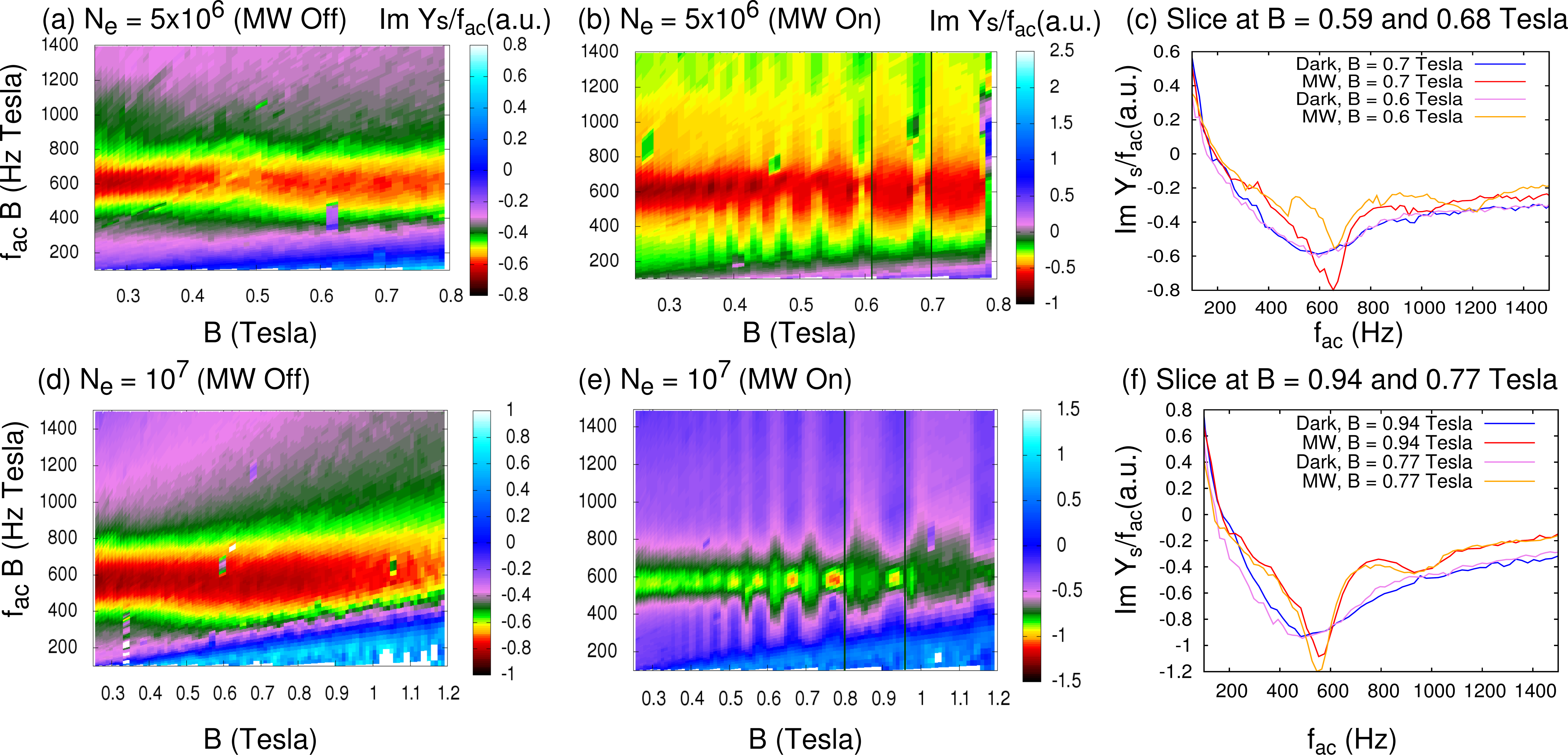}
\caption{\label{figAppendixRF1} The color-scale diagrams show the imaginary part of the admittance from one segment (see Fig.~\ref{figAppendixSegs} for a discussion on the phase of the signal) measured as function of the perpendicular magnetic field $B$ and the low frequency AC excitation frequency $f_{ac}$. In the figure this frequency is multiplied by $B$ to emphasize the $1/B$ dependence of the GMP resonance, indeed with this rescaling the position of the resonance remains fixed as the magnetic field is varied. Experiments are shown both in the dark and under a microwave irradiation at $139$ GHz a frequency which excites the transition between the two lowest lying Rydberg levels leading to microwave induced resistance oscillations which appear in this figures as a $1/B$ periodic modulation of the GMP resonance quality factor. As shown in the data slices on the right panel the quality factor of the GMP resonance substantially increases at minima of microwave induced resistance oscillations. Top figures correspond to $N_e = 5\times 10^6$ while bottom bottom figures were obtained for $N_e = 10\times 10^6$. In both cases the gate voltages where $V_d = 4.5$ and $V_b = -3.5$ Volt.
}
\end{figure*}

 While we do not yet have a model for the resonant modes in the zero resistance state it seems very likely that excitation of the GMP mode plays an important role in the transition to zero-resistance state ZRS. Experimentally self-oscillations at frequencies consistent with the GMP mode were reported in \cite{DenisWatanabe}. We also showed that ZRS for electrons on helium is a collective state where the microwave excitation must be at resonance with the transition between the lowest Rydberg levels in all the system (intersubband resonance), detuning the edge of the electron cloud is enough to suppress ZRS in the center of the cloud even if the edge is separated from the center by a macroscopic distance larger than electrostatic screening length \cite{Alexei2015}. GMP is a collective mode spreading through all the electron cloud and seems well placed to explain this collective behavior. Note that in these experiments anomalous compressibility values (negative) were found close to $V_b = 0$, a value that we now understand to be singular for the resonance frequency of GMP. Finally we also know that due to in-homogeneous broadening only a small fraction of the electrons in the system should actually be at resonance with the microwave photons, yet the collective response seems to suggest that all the electrons are excited. The GMP mode can also provide an explanation to this apparently conflicting observations. Indeed the density oscillations in the GMP mode will also create an oscillation of the perpendicular electric field. If the amplitude of the GMP oscillation is sufficiently large so that it can locally compensate the detuning due to the static inhomogeneous broadening. In this case during a cycle of the GMP mode all the electrons will cross the resonance and will all be excited at least once in a GMP oscillation period. These argumemnts show that an accurate modeling of the GMP resonance is a an important step for the understanding of zero-resistance states for electrons on helium.



\end{document}